# Cold Kaons from Hot Fireballs [*]

Volker Koch [a]

[a]Nuclear Science Division, Lawrence Berkeley Laboratory,
Berkeley, CA 94720, USA

The E814-collaboration has found a component of very low $m_t$ $K^+$ mesons with a slope parameter of $T \sim 15\,\mathrm{MeV}$. We will present a scenario which explains the observed slope parameter and which allows us to predict the expected slope parameter for kaons produced in heavier systems such as Au+Au. We also will discuss the effect of the coulomb interaction on the structure of the spectrum.

## 1. Introduction

At Quark Matter '93 J. Stachel [1] presented the first data on kaon spectra obtained by the E-814 spectrometer at the AGS. These data cover only the very lowest range in $m_t$ ($m_t - m_k \leq 25\,\mathrm{MeV}$) at somewhat forward rapidities ($2.2 \leq Y \leq 2.6$) and they can be parameterized by an exponential in $m_t$ with a slope parameter as low as $T \simeq 15\,\mathrm{MeV}$. In comparison, the E-802 collaboration [2] gives a slope-parameter of $\sim 150\,\mathrm{MeV}$ at central rapidities for kaons with $m_t - m > 100\,\mathrm{MeV}$.

Since it seems very unlikely, that resonance decay can account for such a small slope parameter, in this contribution we will discuss the possibility, that an attractive, long range (mean field) potential may be responsible for the observed enhancement. As has been discussed in detail in [6], in this scenario the slope at small $p_t$ is directly related to the expansion/life time of the system. The measured very small slope parameter, is consistent with a small expansion velocity of $v_{exp} \simeq 0.1c$, about 1/3 of the velocity obtained for cascade type expansion. Thus, one may be lead to speculate, that the system has just crossed the chiral restoration point, where the pressure, which drives the expansion, is small.

Another ingredient of our model is an attractive potential for the $K^+$. An attractive kaon potential would be conceivable in the framework of chiral perturbation theory [4], if one allows for a considerable drop in the vector meson coupling above $T_c$, as suggested by lattice gauge calculations [5]. Here, however, we want to follow a phenomenological approach and determine the strength of the $K^+$ optical potential from the available data.

[*]This work was supported by the Director, Office of Energy Research, Office of High Energy and Nuclear Physics Division of the Department of Energy, under Contract No. DE-AC03-76SF00098.

## 2. Model calculation

A realistic calculation requires to include not only the effect of slow expansion but also the rescattering of the kaons in the medium. Both can be included in a transport calculation, where the slowing down of the expansion can be modeled by a selfconsistent mean field, which in the infinite matter limit exhibits a phase transition. We will use a relativistic transport model (see e.g. [7]), which includes nucleons, deltas, pions and kaons. The collisions among these particles are controlled by standard cross sections [7,8]. The mean field is based on the Walecka- $(\sigma - \omega)$ model, which shows a very rapid rise in the energy density to the value of free, massless particles at a certain critical temperature, which depends on the parameters used [9]. In this model, the necessary field (or decondensation) energy/pressure associated with a change of the gluon condensate at the chiral restoration transition can be simulated by the field energy of the scalar sigma-meson. Since we are only concerned about the expansion dynamics, such an effective approach should be sufficient. For the kaons, we take a potential of the form

$$U_K(r) = U_0 \frac{\rho_s}{\rho_0} \qquad (1)$$

where the density $\rho_s$ is the scalar density $<\bar{\psi}\psi>$.

In Fig. 1a we show our result for the kaon production for Si+Au. Together with the preliminary data, we also show the prediction of the RQMD model [3] determined by the E-814 collaboration. In order to obtain reliable statistics, in our calculation the kaon spectrum has been obtained after averaging over rapidity, while the data are taken at somewhat forward rapidities. The data are reproduced by a kaon potential of $U_0 \simeq -50$ MeV and Walecka model parameters of $g_v = 5.5$ and $g_s = 9.27$. Notice that a smaller vector coupling than the original Walecka value ($g_s^{orig} = 9.27$) is needed. In Fig. 1a we also show the resulting kaon spectrum obtained in the pure cascade approach. Clearly no soft component has developed in this case.

Our prediction for the kaon spectrum in case of Au+Au is shown in Fig. 1b. We simulate the Au+Au collision by assuming a fireball of radius 6.5 fm and with 350 baryons inside the fireball. Again the spectrum develops a concave shape but at low $m_\perp$ the slope is not as steep as that for the smaller system, mainly because of the increased rescattering of the kaons in the bigger fireball.

After the presentation of these results, C.Y. Wong suggested that the coulomb interaction, due to its long range, could lead to a drop in the spectrum at low $p_t$. But also for the coulomb potential, the expansion velocity of the fireball, which carries the charge, is crucial. If the fireball expands rapidly, the momenta of slow particle will not have changed much before the potential has disappeared. To demonstrate this more quantitatively, we have extended our schematic model [6] to include the coulomb force. In this model, the kaons move in a time dependent, expanding potential ($U = U_0 + U_{coulomb}$). Its parameters are the expansion time $\tau$, the initial depth of the potential $U_0$ as well as the initial radius $R_0$ and the charge $Z$ of the fireball. To simulate the system Au + Au we have chosen $R_0 = 6$ MeV and $Z = 150$. We further have assumed $U_0 = -100$ MeV, in order to have an initial net attraction of $-50$ MeV, which was needed in the case without coulomb force [6]. In Fig. 2 we show the resulting kaon spectra for a cylindrical fireball for different expansion times, where longitudinal and transverse expansion times are as-sumed to be equal. In Fig. 2a we show the resulting kaon potential for expansion times $\tau = 3$ fm and $\tau = 10$ fm with and without the attractive potential $U_0$. The drop in the spectrum due to the coulomb interaction is most prominent for the long expansion time. In our favorite scenario (long expansion time, attractive kaon potential (dashed-dotted line)) the predicted enhancement persists for transverse energies above $\simeq 15$ MeV. Below $\sim 10$ MeV, the spectrum falls rapidly by about a factor of two as a result of the coulomb interaction[2]. In the 'conservative' scenario (fast expansion, no kaon potential (dotted line) ) the coulomb force just manages to flatten the slope somewhat, but does not generate a drop at low $m_t$. Incidentally, the position, where the 'coulomb drop' sets in, seems to be related to the depth of the kaon potential while the slope of the drop seems to to be related to the expansion time. It would be interesting to check the latter correlation in case of pion and proton spectra.

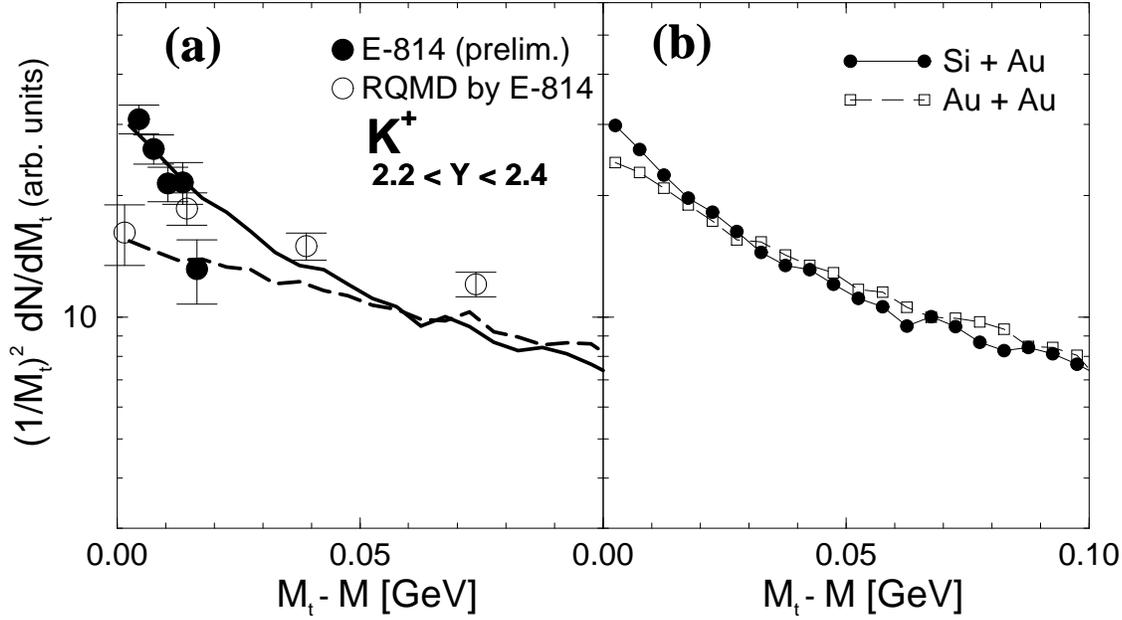

Figure 1. Kaon spectra from transport calculation. (a) Si + Au: Full line is result including mean fields for baryons and kaons. The dashed line is the the pure cascade result. (b) Au+Au.

---

[2]At this point we should point out that the slow expansion has to continue as long as 200 fm/c for the drop in the spectrum to occur. It is very questionable if this will be consistent with the reported observation of radial flow.

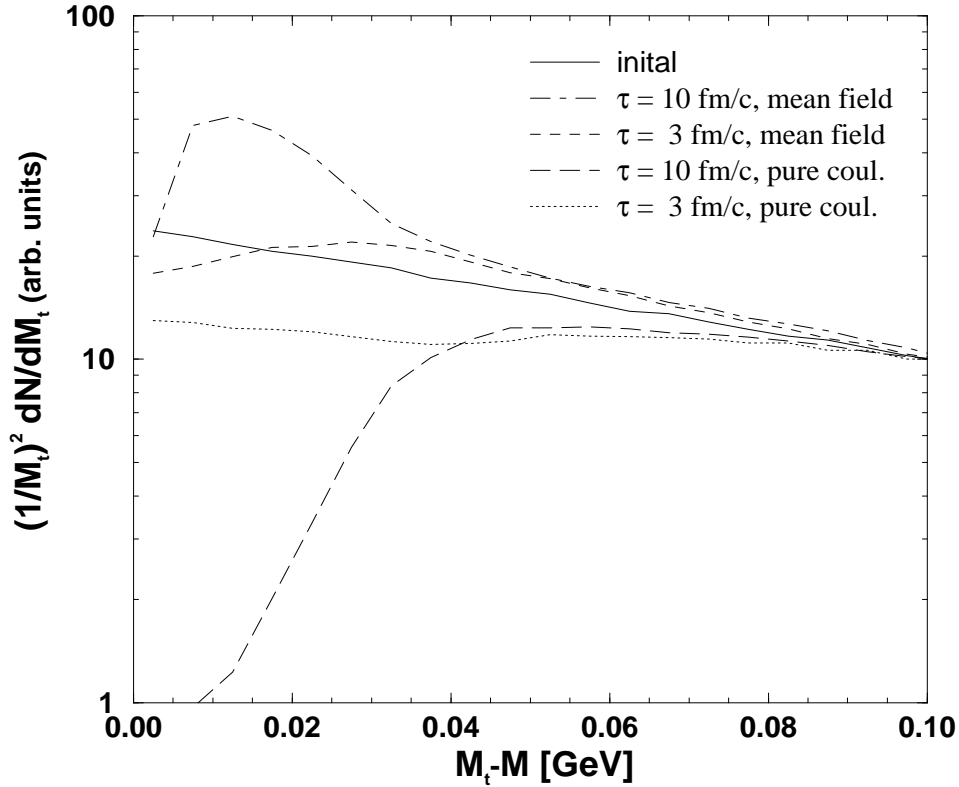

Figure 2. Initial and final kaon spectra for different expansion times $\tau$ including the coulomb interaction. The initial value for the kaon mean field has been taken to be $U_0 = -100\,\text{MeV}$. All spectra are normalized at $m_t - m = 100\,\text{MeV}$.

In conclusion, we have presented a scenario which explains the low $p_t$ enhancement of kaons as seen in preliminary data of the E-814 collaboration. The essential model assumption is the slow expansion of the fireball created in the heavy ion collisions, which may be related to the onset of the chiral restoration transition. We predict less an enhancement for the heavier system $Au + Au$ due to an enhanced rescattering of the kaons in the hot fireball. Including the coulomb force, our slow expansion scenario still predicts an enhancement in the spectrum at small transverse energies, with a sharp drop below $\sim 10\,\text{MeV}$. It also seems, that the slope of the 'coulomb drop' provides another measure for the expansion time while its position may be related to the strength of the attractive kaon potential. For negative kaons, on the other hand, we expect an even steeper rise at low $m_t$. Finally, we predict a disappearance of the cold component with decreasing *as well as* increasing bombarding energy. Below $T_c$ the vector coupling has essentially its zero temperature value and, thus, the kaons should not feel an attractive potential. At higher energies/temperature, the pressure approaches that of an ideal gas, which again leads to a fast expansion of the system.

**Acknowledgements:**

I would like to thank G. Brown for many interesting discussions and C.Y. Wong for pointing out the importance of the coulomb interaction at low $m_t$.